\documentclass[aoas,MSNbibl,nameyear,seceqn,dvips]{arximspdf}
\usepackage{graphicx}

\doi{10.1214/12-AOAS575} 
\volume{7}
\issue{1}
\pubyear{2013}
\firstpage{226}
\lastpage{248}

\makeatletter

\newcommand{\eqref}[1]{(\ref{#1})}
\newcommand{\mymat}[1]{\mathbf{#1}} 
\newcommand{\myvec}[1]{\mathbf{#1}} 
\newcommand{\myobs}[1]{\mathbf{#1}} 
\newcommand{\pkg}[1]{\textit{#1}}
\newcommand{\proglang}[1]{\textsf{#1}}
\newtheorem{theorem}{Theorem}

\makeatother

\begin{document}
\begin{frontmatter}

\title{Sparse least trimmed squares regression for analyzing high-dimensional large data sets}
\runtitle{Sparse least trimmed squares regression}

\begin{aug}
\author[A]{\fnms{Andreas} \snm{Alfons}\ead[label=e1]{andreas.alfons@econ.kuleuven.be}},
\author[A]{\fnms{Christophe} \snm{Croux}\corref{}\ead[label=e2]{christophe.croux@econ.kuleuven.be}}
\and
\author[B]{\fnms{Sarah} \snm{Gelper}\ead[label=e3]{sgelper@rsm.nl}}
\runauthor{A. Alfons, C. Croux and S. Gelper}
\affiliation{KU Leuven, KU Leuven and Erasmus University Rotterdam}
\address[A]{A. Alfons\\
C. Croux \\
ORSTAT Research Center\\
Faculty of Business and Economics\\
KU Leuven\\
Naamsestraat 69\\
3000 Leuven\\
Belgium\\
\printead{e1}\\
\phantom{E-mail:\ }\printead*{e2}}
\address[B]{S. Gelper\\
Rotterdam School of Management\\
Erasmus University Rotterdam\\
Burgemeester Oudlaan 50\\
3000 Rotterdam\\
The Netherlands\\
\printead{e3}}

\end{aug}

\received{\smonth{7} \syear{2011}}
\revised{\smonth{5} \syear{2012}}

%
\begin{abstract}
Sparse model estimation is a topic of high importance in modern data analysis
due to the increasing availability of data sets with a large number of
variables. Another common problem in applied statistics is the presence of
outliers in the data. This paper combines robust regression and sparse model
estimation. A robust and sparse estimator is introduced by adding an $L_{1}$
penalty on the coefficient estimates to the well-known least trimmed squares
(LTS) estimator. The breakdown point of this sparse LTS estimator is derived,
and a fast algorithm for its computation is proposed. In addition, the sparse
LTS is applied to protein and gene expression data of the NCI-60 cancer cell
panel. Both a simulation study and the real data application show that the
sparse LTS has better prediction performance than its competitors in the
presence of leverage points.
\end{abstract}

%

\end{frontmatter}

\section{Introduction}
\label{secintro}

In applied data analysis, there is an increasing availability of data sets
containing a large number of variables. Linear models that include the
full set
of explanatory variables often have poor prediction performance as they tend
to have large variance. Furthermore, large models are in general
difficult to interpret. In many cases, the number of variables is even larger
than the number of observations. Traditional methods such as least
squares can
then no longer be applied due to the rank deficiency of the design matrix.
For instance, gene expression or fMRI studies typically contain tens of
thousands of variables for only a small number of observations.
In this paper, we present an application to the cancer cell panel of the
National Cancer Institute, in which the data consists of $59$
observations and
$22\mbox{,} 283$ predictors.

To improve prediction accuracy and as a remedy to computational
problems with
high-dimensional data, a penalty term on the regression coefficients
can be
added to the objective function. This approach shrinks the coefficients and
reduces variance at the price of increased bias. \citet{tibshirani96}
introduced the least absolute shrinkage and selection operator (lasso),
in which
the penalty function is the $L_{1}$ norm. Let $\myvec{y} = (y_{1},
\ldots,
y_{n})'$ be the response and $\mymat{X} = (x_{ij})_{1 \leq i \leq n, 1
\leq j
\leq p}$ the matrix of predictor variables, where $n$ denotes the
number of
observations and $p$ the number of variables. In addition, let $\myobs{x}_{1},
\ldots, \myobs{x}_{n}$ be the $p$-dimensional observations, that is,
the rows of
$\mymat{X}$. We assume a standard regression model
%
\begin{equation}
\label{mymodel} y_i=\myobs{x}_{i}' \bolds{
\beta} + \varepsilon_i,
\end{equation}
where the regression parameter is $\bolds{\beta} = (\beta_{1},
\ldots,
\beta_{p})'$, and the error terms $\varepsilon_i$ have zero expected value.
With a penalty parameter $\lambda$, the lasso
estimate of $\bolds{\beta}$ is
%
\begin{equation}
\label{eqlasso} \hat{\bolds{\beta}}_{\mathrm{lasso}} = \mathop{\operatorname
{argmin}}_{\bolds{\beta}} \sum_{i = 1}^{n}
\bigl(y_{i} - \myobs{x}_{i}' \bolds{\beta}
\bigr)^{2} + n \lambda \sum_{j=1}^{p}
|\beta_{j}|.
\end{equation}
%
The lasso is frequently used in practice since the $L_{1}$ penalty allows
to shrink some coefficients to exactly zero, that is, to produce sparse model
estimates that are highly interpretable.
In addition, a fast algorithm for
computing the lasso is available through the framework of least angle regression
[LARS; \citet{efron04}]. Other algorithms are available as well
[e.g., \citet{wu08}]. Due to the popularity of the lasso, its theoretical
properties are well studied in the literature [e.g., \citet{knight00},
\citet{zhao06},
\citet{zou07}] and several modifications have been proposed [e.g.,
\citet{zou06},
\citet{yuan06},
\citet{gertheiss10},
\citet{radchenko11},
\citet{wang11}]. However, the lasso is not
robust to
outliers. In this paper we formally show that the breakdown point of
the lasso
is $1/n$, that is, only one single outlier can make the lasso estimate
completely
unreliable. Therefore, robust alternatives are needed.

Outliers are observations that deviate from the model assumptions and
are a
common problem in the practice of data analysis.
For example, for many of the $22\mbox{,}283$ predictors in the NCI data set
used in
Section~\ref{secex}, (log-transformed) responses on the $59$ cell
lines showed
outliers.
Robust alternatives to the least squares regression estimator are well
known and
studied; see \citet{maronna06} for an overview. In this paper, we
focus on the
least trimmed squares (LTS) estimator introduced by \citet{rousseeuw84}.
This estimator has a simple definition, is quite fast to compute, and is
probably the most popular robust regression estimator. Denote the
vector of
squared residuals by $\myvec{r}^{2}(\bolds{\beta}) = (r_{1}^{2},
\ldots,
r_{n}^{2})'$ with $r_{i}^{2} = (y_{i} - \myobs{x}_{i}' \bolds{\beta
})^{2}$, $i =
1, \ldots, n$. Then the LTS estimator is defined as
%
\begin{equation}
\label{eqLTS} \hat{\bolds{\beta}}_{\mathrm{LTS}} = \mathop{\operatorname
{argmin}}_{\bolds{\beta}} \sum_{i = 1}^{h}
\bigl( \myvec{r}^{2}(\bolds{\beta}) \bigr)_{i:n},
\end{equation}
where $(\myvec{r}^{2}(\bolds{\beta}))_{1:n} \leq\cdots\leq
(\myvec{r}^{2}(\bolds{\beta}))_{n:n}$ are the order statistics of
the squared
residuals and $h \leq n$. Thus, LTS regression corresponds to finding
the subset
of $h$ observations whose least squares fit produces the smallest sum
of squared
residuals. The subset size $h$ can be seen as an initial guess of the
number of
good observations in the data. While the LTS is highly robust, it
clearly does
not produce sparse model estimates. Furthermore, if $h<p$, the LTS estimator
cannot be computed. A sparse and regularized version of the LTS is
obtained by
adding an $L_{1}$ penalty with penalty parameter $\lambda$ to~(\ref{eqLTS}),
leading to the \textit{sparse LTS} estimator
%
\begin{equation}
\label{eqsparseLTS} \hat{\bolds{\beta}}_{\mathrm{sparseLTS}} = \mathop{
\operatorname{argmin}}_{\bolds{\beta}} \sum_{i = 1}^{h}
\bigl( \myvec{r}^{2}(\bolds{\beta}) \bigr)_{i:n} + h \lambda
\sum_{j = 1}^{p} |\beta_{j}|.
\end{equation}
We prove in this paper that sparse LTS has a high breakdown point. It is
resistant to multiple regression outliers, including leverage points. Besides
being highly robust, and similar to the lasso estimate, sparse LTS (i) improves
the prediction performance through variance reduction if the sample
size is
small relative to the dimension, (ii) ensures higher interpretability
due to
simultaneous model selection, and (iii) avoids computational problems of
traditional robust regression methods in the case of high-dimensional data.
For the NCI data, sparse LTS was less influenced by the outliers than competitor
methods and showed better prediction performance, while the resulting
model is
small enough to be easily interpreted (see Section~\ref{secex}).

The sparse LTS (\ref{eqsparseLTS}) can also be interpreted as a trimmed
version of the lasso, since the limit case $h=n$ yields the lasso solution.
Other robust versions of the lasso have been considered in the
literature. Most
of them are penalized M-estimators, as in \citet{vandegeer08} and
\citet{li11}.
\citet{rosset04} proposed a Huber-type loss function, which requires knowledge
of the residual scale. A least absolute deviations (LAD) type of estimator
called LAD-lasso is proposed by \citet{wang07},
%
\begin{equation}
\label{LAD-lasso} \hat{\bolds{\beta}}_{\mathrm{LAD\mbox{-}lasso}} = \mathop{
\operatorname{argmin}}_{\bolds{\beta}} \sum_{i=1}^{n}
\bigl|y_{i} - \myobs{x}_{i}' \bolds{\beta}\bigr| + n
\lambda\sum_{j=1}^{p} |\beta_{j}|.
\end{equation}
However, none of these methods is robust with respect to leverage
points, that is,
outliers in the predictor space, and can handle outliers only in the response
variable. The main competitor of the sparse LTS is robust least angle
regression, called RLARS, and proposed in \citet{khan07b}. They
develop a robust
version of the LARS algorithm, essentially replacing correlations by a robust
type of correlation, to sequence and select the most important predictor
variables. Then a nonsparse robust regression estimator is applied to the
selected predictor variables. RLARS, as will be confirmed by our simulation\vadjust{\goodbreak}
study, is robust with respect to leverage points. A main drawback of
the RLARS
algorithm of \citet{khan07b} is the lack of a natural definition,
since it is
not optimizing a clearly defined objective function.

An entirely different approach is taken by \citet{she11}, who propose an
iterative procedure for outlier detection. Their method is based on
imposing a
sparsity criterion on the estimator of the mean-shift parameter $\bolds{\gamma}$
in the extended regression model
%
\begin{equation}
\myvec{y} = \mymat{X} \bolds{\beta} + \bolds{\gamma} + \bolds{\varepsilon}.
\end{equation}
They stress that this method requires a nonconvex sparsity criterion. An
extension of the method to high-dimensional data is obtained by also assuming
sparsity of the coefficients $\bolds{\beta}$. Nevertheless, their
paper mainly
focuses on outlier detection and much less on sparse robust estimation. Note
that another procedure for simultaneous outlier identification and variable
selection based on the mean-shift model is proposed by
\citet{menjoge10}.\looseness=-1

The rest of the paper is organized as follows. In Section~\ref
{secfbp} the
breakdown point of the sparse LTS estimator is obtained. Further, we
also show
that the lasso and the LAD-lasso have a breakdown point of only $1/n$. A
detailed description of the proposed algorithm to compute the sparse LTS
regression estimator is provided in Section~\ref{secalg}.
Section~\ref{secreweighted} introduces a reweighted version of the
estimator in
order to increase statistical efficiency. The choice of the penalty parameter
$\lambda$ is discussed in Section~\ref{seclambda}. Simulation
studies are
performed in Section~\ref{secsim}. In addition, Section~\ref{secex} presents
an application to protein and gene expression data of the well-known
cancer cell
panel of the National Cancer Institute.
The results indicate that these data contain outliers such that robust methods
are necessary for analysis. Moreover, sparse LTS yields a model that is
easy to
interpret and has excellent prediction performance.
Finally, Section~\ref{seccpu} presents some computation times and
Section~\ref{secconcl} concludes.


\section{Breakdown point}
\label{secfbp}

The most popular measure for the robustness of an estimator is the
\textit{replacement finite-sample breakdown point} (FBP) [e.g., \citet
{maronna06}].
Let $\mymat{Z} = (\mymat{X}, \myvec{y})$ denote the sample. For a regression
estimator $\hat{\bolds{\beta}}$, the breakdown point is defined as
%
\begin{equation}
\varepsilon^{*}(\hat{\bolds{\beta}}; \mymat{Z}) = \min \biggl\{
\frac{m}{n} \dvtx \sup_{\tilde{\mymat Z}} \bigl\| \hat{\bolds{\beta}}(\tilde{\mymat
{Z}}) \bigr\|_{2} = \infty \biggr\},
\end{equation}
where $\tilde{\mymat{Z}}$ are corrupted data obtained from $\mymat
{Z}$ by
replacing $m$ of the original $n$ data points by arbitrary values. We obtained
the following result, from which the breakdown point of the sparse LTS estimator
immediately follows. The proof is in the \hyperref[app]{Appendix}.

%
\begin{theorem} \label{thmfbp}
Let $\rho(x)$ be a convex and symmetric loss function with $\rho(0) =
0$ and
$\rho(x) > 0$ for $ x \neq0$, and define\vadjust{\goodbreak} $\bolds{\rho}(\myvec{x}) :=
(\rho(x_{1}), \ldots, \rho(x_{n}))'$. With subset size $h \leq n$,
consider the
regression estimator
%
\begin{equation}
\label{eqtrimmedLassoType} \hat{\bolds{\beta}} = \mathop{\operatorname{argmin}}_{\bolds{\beta}}
\sum_{i = 1}^{h} \bigl( \bolds{\rho}(\myvec{y}
- \mymat{X} \bolds{\beta}) \bigr)_{i:n} + h \lambda\sum
_{j = 1}^{p} |\beta_{j}|,
\end{equation}
where $ ( \bolds{\rho}(\myvec{y} - \mymat{X} \bolds{\beta})
))_{1:n}
\leq\cdots\leq ( \bolds{\rho}(\myvec{y} - \mymat{X} \bolds{\beta})
)_{n:n}$ are the order statistics of the regression loss. Then the
breakdown point of the estimator $\hat{\bolds{\beta}}$ is given by
\[
\varepsilon^{*}(\hat{\bolds{\beta}}; \mymat{Z}) = \frac{n-h+1}{n}.
\]
\end{theorem}
%

The breakdown point is the same for any loss function $\rho$
fulfilling the
assumptions. In particular, the breakdown point for the sparse LTS estimator
$\hat{\beta}_{\mathrm{sparseLTS}}$ with subset size $h \leq n$, in
which $\rho(x)
= x^{2}$, is still $(n-h+1)/n$.
The smaller the value of $h$, the higher the breakdown point. By taking $h$
small enough, it is even possible to have a breakdown point larger than
50\%.
However, while this is mathematically possible, we are not advising to
use $h <
n/2$ since robust statistics aim for models that fit the majority of
the data.
Thus, we do not envisage to have such large breakdown points. Instead, we
suggest to take a value of $h$ equal to a fraction $\alpha$ of the
sample size,
with $\alpha=0.75$, such that the final estimate is based on a sufficiently
large number of observations. This guarantees a sufficiently high statistical
efficiency, as will be shown in the simulations in Section~\ref
{secsim}. The
resulting breakdown point is then about $1-\alpha=25\%$. Notice that the
breakdown point does not depend on the dimension $p$. Even if the
number of
predictor variables is larger than the sample size, a high breakdown
point is
guaranteed. For the nonsparse LTS, the breakdown point does depend on $p$
[see \citet{rousseeuw03}].

Applying Theorem~\ref{thmfbp} to the lasso [corresponding to $\rho
(x) = x^{2}$
and $h=n$] yields a finite-sample breakdown point of
\[
\varepsilon^*(\hat{\bolds{\beta}}_{\mathrm{lasso}}; \mymat {Z})=\frac{1}{n}.
\]
Hence, only one outlier can already send the lasso solution to
infinity, despite
the fact that large values of the regression estimate are penalized in the
objective function of the lasso. The nonrobustness of the Lasso comes
from the
use of the squared residuals in the objective function \eqref
{eqlasso}. Using
other convex loss functions, as done in the LAD-lasso or penalized M-estimators,
does not solve the problem and results in a breakdown point of $1/n$ as well.
The theoretical results on robustness are also reflected in the
application to
the NCI data in Section~\ref{secex}, where the lasso is much more
influenced by
the outliers than the sparse LTS.


\section{Algorithm}
\label{secalg}

We first present an equivalent formulation of the sparse LTS estimator
(\ref{eqsparseLTS}). For a fixed penalty parameter $\lambda$, define\vadjust{\goodbreak} the
objective function
%
\begin{equation}
\label{eqobj} Q(H, \bolds{\beta}) = \sum_{i \in H}
\bigl(y_{i} - \myobs{x}_{i}' \beta
\bigr)^{2} + h \lambda\sum_{j=1}^{p}
|\beta_{j}|,\vadjust{\goodbreak}
\end{equation}
which is the $L_{1}$ penalized residual sum of squares based on a
subsample $H
\subseteq\{ 1, \ldots, n \}$ with $|H| = h$. With
%
\begin{equation}
\label{lassofith} \hat{\bolds{\beta}}_{H} = \mathop{
\operatorname{argmin}}_{\bolds{\beta}} Q(H, \bolds{\beta}),
\end{equation}
the sparse LTS
estimator is given by $\hat{\bolds{\beta}}_{H_{\mathrm{opt}}}$, where
%
\begin{equation}
\label{eqHopt} H_{\mathrm{opt}} = \mathop{\operatorname{argmin}}_{H \subseteq\{ 1,
\ldots, n \}:|H|
= h} Q(H,
\hat{\bolds{\beta}}_{H}).
\end{equation}
Hence, the sparse LTS corresponds to finding the subset of $h \leq n$
observations whose lasso fit produces the smallest penalized residual
sum of
squares. To find this optimal subset, we use an analogue of the FAST-LTS
algorithm developed by \citet{rousseeuw06}.

The algorithm is based on \textit{concentration steps} or C-steps. The
C-step at
iteration $k$ consists of computing the lasso solution based on the current
subset $H_{k}$, with $|H_{k}| = h$, and constructing the next subset $H_{k+1}$
from the observations corresponding to the $h$ smallest squared
residuals. Let
$H_{k}$ denote a certain subsample derived at iteration $k$ and let
$\hat{\bolds{\beta}}_{H_{k}}$\vspace*{-1pt} be the coefficients of the
corresponding lasso
fit. After computing the squared residuals $\myvec{r}_{k}^{2} = (r_{k,1}^{2},
\ldots, r_{k,n}^{2})'$ with $r_{k,i}^{2} = (y_{i} - \myobs{x}_{i}'
\hat{\bolds{\beta}}_{H_{k}})^{2}$, the subsample $H_{k+1}$ for
iteration $k+1$
is defined as the set of indices corresponding to the $h$ smallest squared
residuals. In mathematical terms, this can be written as
%
\[
H_{k+1} = \bigl\{ i \in\{ 1, \ldots, n \} \dvtx
r_{k,i}^{2} \in\bigl\{ \bigl(\myvec{r}_{k}^{2}
\bigr)_{j:n} \dvtx j = 1, \ldots, h \bigr\} \bigr\},
\]
where $(\myvec{r}_{k}^{2})_{1:n} \leq\cdots\leq(\myvec{r}_{k}^{2})_{n:n}$
denote the order statistics of the squared residuals. Let
$\hat{\bolds{\beta}}_{H_{k+1}}$ denote coefficients of the lasso fit
based on
$H_{k+1}$. Then
%
\begin{equation}
\label{eqdecrease} Q(H_{k+1}, \hat{\bolds{\beta}}_{H_{k+1}}) \leq
Q(H_{k+1}, \hat{\bolds{\beta}}_{H_{k}}) \leq Q(H_{k},
\hat{\bolds{\beta}}_{H_{k}}),
\end{equation}
where the first inequality follows from the definition of
$\hat{\bolds{\beta}}_{H_{k+1}}$, and the second inequality from the
definition
of $H_k$. From \eqref{eqdecrease} it follows that a C-step results in a
decrease of the sparse LTS objective function, and that a sequence of C-steps
yields convergence to a local minimum in a finite number of steps.

To increase the chances of arriving at the global minimum, a
sufficiently large
number $s$ of initial subsamples $H_0$ should be used, each of them
being used
as starting point for a sequence of C-steps. Rather than randomly
selecting $h$
data points, any initial subset $H_{0}$ of size $h$ is constructed from an
\textit{elemental subset} of size 3 as follows. Draw three
observations from the\vadjust{\goodbreak}
data at random, say, $\myobs{x}_{i_{1}}$, $\myobs{x}_{i_{2}}$ and
$\myobs{x}_{i_{3}}$. The lasso fit for this elemental subset is then
%
\begin{equation}
\label{eqinitial} \hat{\bolds{\beta}}_{\{{i_{1}}, {i_{2}},{i_{3}}
\}} = \mathop{
\operatorname{argmin}}_{\bolds{\beta}} Q\bigl(\{{i_{1}}, {i_{2}},{i_{3}}
\}, \bolds{\beta}\bigr),
\end{equation}
and the initial subset $H_{0}$ is then given by the indices of the $h$
observations with the smallest squared residuals with respect to the
fit in
\eqref{eqinitial}. The nonsparse FAST-LTS algorithm uses elemental
subsets of
size $p$, since any OLS regression requires at least as many
observations as the
dimension $p$. This would make the algorithm not applicable if $p > n$.
Fortunately the lasso is already properly defined for samples of size
3, even
for large values of $p$. Moreover, from a robustness point of view,
using only
three observations is optimal, as it ensures the highest probability of not
including outliers in the elemental set. It is important to note that the
elemental subsets of size 3 are only used to construct the initial
subsets of
size $h$ for the C-step algorithms. All C-steps are performed on
subsets of size
$h$.

In this paper, we used $s = 500$ initial subsets.
Using a larger number of subsets did not lead to better prediction performance
in the case of the NCI data.
Following the strategy advised in \citet{rousseeuw06}, we perform only two
C-steps for all $s$ subsets and retain the $s_{1}=10$ subsamples with the
lowest values of the objective function~\eqref{eqobj}. For the
reduced number
of subsets $s_{1}$, further C-steps are performed until convergence.
This is a
standard strategy for C-step algorithms to decrease computation time.

\textit{Estimation of an intercept}: the regression model in
\eqref{mymodel} does not contain an intercept. It is indeed common to assume
that the variables are mean-centered and the predictor variables are
standardized before applying the lasso. However, computing the means and
standard deviations over all observations does not result in a robust
method, so
we take a different approach. Each time the sparse LTS algorithm
computes a
lasso fit on a subsample of size $h$, the variables are first centered
and the
predictors are standardized using the means and standard deviations computed
from the respective subsample. The resulting procedure then minimizes
\eqref{eqsparseLTS} with squared residuals $r_{i}^{2} = (y_{i} -
\beta_{0} -
\myobs{x}_{i}' \bolds{\beta})^{2}$, where $\beta_{0}$ stands for
the intercept.
We verified that adding an intercept to the model has no impact on the breakdown
point of the sparse LTS estimator of $\bolds{\beta}$.


\section{Reweighted sparse LTS estimator}
\label{secreweighted}

Let $\alpha$ denote the proportion of observations from the full
sample to be
retained in each subsample, that is, $h = \lfloor(n + 1) \alpha
\rfloor$. In this
paper we take $\alpha= 0.75$. Then $(1-\alpha)$ may be interpreted as an
initial guess of the proportion of outliers in the data. This initial guess
is typically rather conservative to ensure that outliers do not impact the
results, and may therefore result in a loss of statistical efficiency. To
increase efficiency, a reweighting step that downweights outliers
detected by
the sparse LTS estimator can be performed.\vadjust{\goodbreak}

Under the normal error model, observations with standardized residuals larger
than a certain quantile of the standard normal distribution may be
declared as
outliers. Since the sparse LTS estimator---like the lasso---is
biased, we need
to center the residuals. A natural estimate for the center of the
residuals is
%
\begin{equation}
\hat{\mu}_{\mathrm{raw}} = \frac{1}{h} \sum_{i \in H_{\mathrm
{opt}}}
r_{i},
\end{equation}
where $r_{i} = y_{i} - \myobs{x}_{i}' \hat{\bolds{\beta}}_{\mathrm
{sparseLTS}}$
and $H_{\mathrm{opt}}$ is the optimal subset from \eqref{eqHopt}.
Then the
residual scale estimate associated to the raw sparse LTS estimator is
given by
%
\begin{equation}
\label{eqrawScale} \hat{\sigma}_{\mathrm{raw}} = k_{\alpha} \sqrt
{\frac{1}{h} \sum_{i=1}^{h}
\bigl( \myvec{r}_{\mathrm{c}}^{2} \bigr)_{i:n}},
\end{equation}
with squared centered residuals $\myvec{r}_{\mathrm{c}}^{2} =  (
(r_{1} -
\hat{\mu}_{\mathrm{raw}})^{2}, \ldots, (r_{n} - \hat{\mu
}_{\mathrm{raw}})^{2}
)'$,
and
%
\begin{equation}
\label{eqconsistency} k_{\alpha} = \biggl( \frac{1}{\alpha} \int
_{-\Phi^{-1}((\alpha+1)/2)}^{\Phi^{-1}((\alpha+1)/2)} u^{2} \,d\Phi(u)
\biggr)^{-1/2},
\end{equation}
a factor to ensure that $\hat{\sigma}_{\mathrm{raw}}$ is a
consistent estimate of
the standard deviation at the normal model. This formulation allows to define
binary weights
%
\begin{equation}
\label{defw} \quad w_{i} = \cases{
1, &\quad
$\mbox{if } \bigl|(r_{i} - \hat{\mu}_{\mathrm{raw}}) / \hat{
\sigma}_{\mathrm{raw}}\bigr| \leq\Phi^{-1}(1-\delta),$
\vspace*{2pt}\cr
0, &\quad $\mbox{if }\bigl |(r_{i} - \hat{\mu}_{\mathrm{raw}}) / \hat{
\sigma}_{\mathrm{raw}}\bigr| > \Phi^{-1}(1-\delta),$}\qquad
i = 1, \ldots, n.
\end{equation}
In this paper $\delta= 0.0125$ is used such that 2.5\% of the
observations are
expected to be flagged as outliers in the normal model, which is a typical
choice.\looseness=-1

The \textit{reweighted sparse LTS} estimator is given by the weighted
lasso fit
%
\begin{equation}
\hat{\bolds{\beta}}_{\mathrm{reweighted}} = \mathop{\operatorname{argmin}}_{\bolds{\beta}}
\sum_{i = 1}^{n} w_{i}
\bigl(y_{i} - \myobs{x}_{i}' \bolds{\beta}
\bigr)^{2} + \lambda n_w \sum_{j =
1}^{p}
|\beta_{j}|,
\end{equation}
with $n_w= \sum_{i=1}^{n} w_{i}$ the sum of weights. With the choice
of weights
given in \eqref{defw}, the reweighted sparse LTS is the lasso fit
based on the
observations not flagged as outliers. Of course, other weighting
schemes could
be considered. Using the residual center estimate
%
\begin{equation}
\hat{\mu}_{\mathrm{reweighted}} = \frac{1}{n_w} \sum_{i=1}^{n}
w_{i} \bigl( y_{i} - \myobs{x}_{i}'
\hat{\bolds{\beta}}_{\mathrm{reweighted}} \bigr),
\end{equation}
the residual scale estimate of the reweighted sparse LTS estimator is
given~by
%
\begin{equation}
\label{eqreweightedScale} \hat{\sigma}_{\mathrm{reweighted}} = k_{\alpha_{w}} \sqrt
{\frac{1}{n_w} \sum_{i=1}^{n}
w_{i} \bigl( y_{i} - \myobs{x}_{i}'
\hat{\bolds{\beta}}_{\mathrm{reweighted}} - \hat{\mu}_{\mathrm
{reweighted}}
\bigr)^{2}},
\end{equation}
where $k_{\alpha_{w}}$ is the consistency factor from (\ref{eqconsistency})
with $\alpha_{w} = n_w/n$.\vadjust{\goodbreak}

Note that this reweighting step is conceptually different from the adaptive
lasso by \citet{zou06}. While the adaptive lasso derives individual
penalties on
the predictors from initial coefficient estimates, the reweighted
sparse LTS
aims to include all nonoutlying observations into fitting the model.


\section{Choice of the penalty parameter}
\label{seclambda}
In practical data analysis, a suitable value of the penalty parameter
$\lambda$
is not known in advance. We propose to select $\lambda$ by optimizing
the Bayes
Information Criterion (BIC), or the estimated prediction performance via
cross-validation. In this paper we use the BIC since it requires less
computational effort. The BIC of a given model estimated with shrinkage
parameter $\lambda$ is given by
%
\begin{equation}
\label{eqBIC} \operatorname{BIC}(\lambda) = \log(\hat{\sigma}) + \mathit{df}(\lambda)
\frac{\log(n)}{n},
\end{equation}
where $\hat{\sigma}$ denotes the corresponding residual scale estimate,
(\ref{eqrawScale}) or (\ref{eqreweightedScale}), and $\mathit{df}(\lambda)$
are the
degrees of freedom of the model. The degrees of freedom are given by
the number
of nonzero estimated parameters in $\hat{\bolds{\beta}}$ [see \citet{zou07}].

As an alternative to the BIC, cross-validation can be used. To prevent outliers
from affecting the choice of $\lambda$, a robust prediction loss
function should
be used. A natural choice is the root trimmed mean squared prediction error
(RTMSPE) with the same trimming proportion as for computing the sparse LTS.
In $k$-fold cross-validation, the data are split randomly in $k$ blocks of
approximately equal size. Each block is left out once to fit the model,
and the
left-out block is used as test data. In this manner, and for a given
value of
$\lambda$, a prediction is obtained for each observation in the
sample. Denote
the vector of squared prediction errors ${\myvec{e}}^2 = (e_{1}^{2},
\ldots,
e_{n}^{2})'$. Then
%
\begin{equation}
\label{RTMSPE} \operatorname{RTMSPE}(\lambda) = \sqrt{\frac{1}{h} \sum
_{i=1}^{h} \bigl({\myvec{e}}^2
\bigr)_{i:n}}.
\end{equation}
To reduce variability, the RTMSPE may be averaged over a number of different
random splits of the data.

The selected $\lambda$ then minimizes $\operatorname{BIC}(\lambda)$ or
$\operatorname{RTMSPE}(\lambda)$ over a grid of values in the interval
$[0, \hat{\lambda}_0]$. We take a grid with steps of size $0.025
\hat{\lambda}_0$, where $\hat{\lambda}_0$ is an estimate of the shrinkage
parameter $\lambda_{0}$ that would shrink all parameters to
zero. If $p > n$, 0 is of course excluded from the grid. For the lasso solution
we take
%
\begin{equation}
\label{eqlambda0} \hat{\lambda}_{0} = \frac{2}{n}
\max_{j \in\{ 1, \ldots, p \}} \operatorname{Cor}(\myvec{y}, \myvec{x}_{j}),
\end{equation}
exactly the same as given and motivated in \citet{efron04}.
In~\eqref{eqlambda0}, $\operatorname{Cor}(\myvec{y}, \myvec{x}_{j})$
stands for the
Pearson correlation between $\myvec{y}$ and the $j$th column of the design
matrix $\mymat{X}$. For sparse LTS, we need a robust estimate
$\hat{\lambda}_{0}$. We propose to replace the Pearson correlation in
\eqref{eqlambda0} by the robust correlation based on bivariate
winsorization of the data [see \citet{khan07b}].


\section{Simulation study}
\label{secsim}

This section presents a simulation study for comparing the performance of
various sparse estimators. The simulations are performed in \proglang{R}
[\citet{RDev}] with package \pkg{simFrame} [\citet{alfons10d},
\citet{simFrame}], which is a
general framework for simulation studies in statistics. Sparse LTS is evaluated
for the subset size $h = \lfloor(n+1) 0.75 \rfloor$. Both the raw and the
reweighted version (see Section~\ref{secreweighted}) are considered.
We prefer
to take a relatively large trimming proportion to guarantee a breakdown
point of
25\%. Adding the reweighting step will then increase the statistical efficiency
of sparse LTS. We make a comparison with the lasso, the LAD-lasso and robust
least angle regression (RLARS), discussed in the introduction. We
selected the
LAD-lasso estimator as a representative of the class of penalized M-estimators,
since it does not need an initial residual scale estimator.

For every generated sample, an optimal value of the shrinkage parameter
$\lambda$ is selected. The penalty parameters for sparse LTS and the
lasso are
chosen using the BIC, as described in Section~\ref{seclambda}. For the
LAD-lasso, we estimate the shrinkage parameter in the same way as in
\citet{wang07}. However, if $p > n$, we cannot use their approach and
use the
BIC as in \eqref{eqBIC}, with the mean absolute value of residuals (multiplied
by a consistency factor) as scale estimate. For RLARS, we add the sequenced
variables to the model in a stepwise fashion and fit robust MM-regressions
[\citet{yohai87}], as advocated in \citet{khan07b}. The optimal model
when using
RLARS is then again selected via BIC, now using the robust scale estimate
resulting from the MM-regression.

\subsection{Sampling schemes}
\label{secsim-settings}

The first configuration is a latent factor model taken from \citet{khan07b}
and covers the case of $n > p$. From $k = 6$ latent independent
standard normal
variables $\myvec{l}_{1}, \ldots, \myvec{l}_{k}$ and an independent
normal error variable $\myvec{e}$ with standard deviation $\sigma$,
the response variable $\myvec{y}$ is constructed as
%
\[
\myvec{y} := \myvec{l}_{1} + \cdots+ \myvec{l}_{k}
+ \myvec{e},
\]
where $\sigma$ is chosen so that the signal-to-noise ratio is 3, that is,
$\sigma=\sqrt{k}/3.$ With independent standard normal variables
$\myvec{e}_{1},
\ldots, \myvec{e}_{p}$, a set of $p = 50$ candidate predictors is then
constructed as
%
\begin{eqnarray*}
\myvec{x}_{j} & := &
\myvec{l}_{j} + \tau\myvec{e}_{j},\qquad j = 1, \ldots, k,
\\
\myvec{x}_{k+1} & := & \myvec{l}_{1} +
\delta\myvec{e}_{k+1},
\\
\myvec{x}_{k+2} & := & \myvec{l}_{1} + \delta
\myvec{e}_{k+2},
\\
& \vdots&
\\
\myvec{x}_{3k-1} & := & \myvec{l}_{k} + \delta
\myvec{e}_{3k-1},
\\
\myvec{x}_{3k} & := & \myvec{l}_{k} + \delta
\myvec{e}_{3k},
\\
\myvec{x}_{j} & := & \myvec{e}_{j},\qquad j =
3k+1, \ldots, p,
\end{eqnarray*}
where $\tau= 0.3$ and $\delta= 5$ so that $\myvec{x}_{1}, \ldots,
\myvec{x}_{k}$ are low-noise perturbations of the latent variables,
$\myvec{x}_{k+1}, \ldots, \myvec{x}_{3k}$ are noise covariates that are
correlated with the latent variables, and $\myvec{x}_{3k+1}, \ldots,
\myvec{x}_{p}$ are independent noise covariates. The number of
observations is
set to $n = 150$.

The second configuration covers the case of moderate high-dimensional
data. We
generate $n = 100$ observations from a $p$-dimensional normal
distribution $N(0,
\Sigma)$, with $p=1 000$. The covariance matrix $\bolds{\Sigma} =
(\Sigma_{ij})_{1 \leq i, j \leq p}$ is given by $\Sigma_{ij} = 0.5^{|i-j|}$,
creating correlated predictor variables. Using the coefficient vector
$\bolds{\beta} = (\beta_{j})_{1 \leq j \leq p}$ with $\beta_{1} =
\beta_{7} =
1.5$, $\beta_{2} = 0.5$, $\beta_{4} = \beta_{11} = 1$, and $\beta_{j} = 0$ for
$j \in\{ 1, \ldots, p \} \setminus\{ 1, 2, 4, 7, 11 \}$, the response
variable is generated according to the regression model \eqref
{mymodel}, where
the error terms follow a normal distribution with $\sigma=0.5$.

Finally, the third configuration represents a more extreme case of
high-dimensional data with $n = 100$ observations and $p = 20\mbox{,}000$ variables.
The first $1 000$ predictor variables are generated from a multivariate normal
distribution $N(0, \Sigma)$ with $\Sigma_{ij} = 0.6^{|i-j|}$.
Furthermore, the
remaining $19\mbox{,}000$ covariates are standard normal variables. Then the response
variable is generated according to \eqref{mymodel}, where the
coefficient vector
$\bolds{\beta}= (\beta_{j})_{1 \leq j \leq p}$ is given by $\beta_{j} = 1$ for
$1 \leq j \leq10$ and $\beta_{j} = 0$ for $11 \leq j \leq p$, and the error
terms follow a standard normal distribution.

For each of the three simulation settings, we apply contamination
schemes taken
from \citet{khan07b}. To be more precise, we consider the following:
\begin{longlist}[(1)]
\item[(1)]\textit{No contamination.}
\item[(2)]\textit{Vertical outliers}: 10\% of the error terms in the regression
model follow a normal $N(20,\sigma)$ instead of a $N(0,\sigma)$.
\item[(3)]\textit{Leverage points}: Same as in 2, but the 10\% contaminated
observations contain high-leverage values by drawing the predictor variables
from independent $N(50, 1)$ distributions.
\end{longlist}
In addition, we investigate a fourth and more stressful outlier scenario.
Keeping the contamination level at 10\%, outliers in the predictor
variables are
drawn from independent $N(10,0.01)$ distributions. Note the small standard
deviation such that the outliers form a dense cluster. Let
$\tilde{\myobs{x}}_{i}$ denote such a leverage point. Then the values
of the
response variable of the contaminated observations are generated by
$\tilde{y}_{i} = \eta\tilde{\myobs{x}}_{i}' \bolds{\gamma}$ with
$\bolds{\gamma} = (-1/p)_{1 \leq j \leq p}$. The direction of $\bolds{\gamma}$
is very different from the one of the true regression parameter $\bolds{\beta}$
in the following ways. First, $\bolds{\gamma}$ is not sparse. Second, all
predictors have a negative effect on the response in the contaminated
observations, whereas the variables with nonzero coefficients have a positive
effect on the response in the good data points. Furthermore, the parameter
$\eta$ controls the magnitude of the leverage effect and is varied
from $1$ to
$25$ in five equidistant steps.

This results in a total of 12 different simulations schemes, which we
think to
be representative for the many different simulation designs we tried out.
The first scheme has $n > p$, the second setting has $p > n$, and the third
setting has $p \gg n$. The choices for the contamination schemes are standard,
inducing both vertical outliers and leverage points in the samples.

\subsection{Performance measures}
Since one of the aims of sparse model estimation is to improve prediction
performance, the different estimators are evaluated by the \textit
{root mean
squared prediction error} (RMSPE). For this purpose, $n$ additional observations
from the respective sampling schemes (without outliers) are generated
as test
data, and this in each simulation run. Then the RMSPE is given by
%
\[
\operatorname{RMSPE}(\hat{\bolds{\beta}}) = \sqrt{
\frac{1}{n} \sum_{i=1}^{n}
\bigl(y^{*}_{i} - \myobs{x}^{*\prime}_{i}
\hat{\bolds{\beta}}\bigr)^{2}},
\]
where $y^{*}_{i}$ and $\myobs{x}^{*}_{i}$, $i = 1, \ldots, n$, denote
the observations of the response and predictor variables in the test data,
respectively. The RMSPE of the oracle estimator, which uses the true coefficient
values $\bolds{\beta}$, is computed as a benchmark for the evaluated
methods. We
report average RMSPE over all simulation runs.

Concerning sparsity, the estimated models are evaluated by the \textit{false
positive rate} (FPR) and the \textit{false negative rate} (FNR). A
false positive
is a coefficient that is zero in the true model, but is estimated as nonzero.
Analogously, a false negative is a coefficient that is nonzero in the true
model, but is estimated as zero. In mathematical terms, the FPR and FNR are
defined as
\begin{eqnarray*}
\operatorname{FPR}(\hat{\bolds{\beta}}) &=& \frac{|\{j \in\{ 1, \ldots,
p \}\dvtx
\hat{\beta}_{j} \neq0 \wedge\beta_{j} = 0 \}|}{|\{j \in\{ 1,
\ldots, p\}\dvtx
\beta_{j} = 0 \}|},
\\
\operatorname{FNR}(\hat{\bolds{\beta}}) &=& \frac{|\{j \in\{ 1,
\ldots, p \}\dvtx
\hat{\beta}_{j} = 0 \wedge\beta_{j} \neq0 \}|}{|\{j \in\{ 1,
\ldots, p \}\dvtx
\beta_{j} \neq0 \}|}.
\end{eqnarray*}
Both FPR and FNR should be as small as possible for a sparse estimator
and are
averaged over all simulation runs. Note that false negatives in general
have a
stronger effect on the RMSPE than false positives. A false negative
means that
important information is not used for prediction, whereas a false positive
merely adds a bit of variance.

\subsection{Simulation results} \label{secsim-results}

In this subsection the simulation results for the different data configurations
are presented and discussed.

%
\begin{table}
\tabcolsep=0pt
\caption{Results for the first simulation scheme, with $n=150$ and
$p=50$. Root
mean squared prediction error (RMSPE), the false positive rate (FPR)
and the
false negative rate (FNR), averaged over 500 simulation runs, are
reported for
every method}\label{tabkhan}
%
\begin{tabular*}{\textwidth}{@{\extracolsep{\fill}}lccccccccc@{}}
\hline& \multicolumn{3}{c}{\textbf{No contamination}} &\multicolumn{3}{c}{\textbf{Vertical outliers}} &\multicolumn{3}{c@{}}{\textbf{Leverage points}}
\\[-6pt]
& \multicolumn{3}{c}{\hrulefill} &\multicolumn{3}{c}{\hrulefill} &\multicolumn{3}{c@{}}{\hrulefill} \\
\textbf{Method} & \textbf{RMSPE} & \textbf{FPR} & \textbf{FNR} & \textbf{RMSPE} & \textbf{FPR} & \textbf{FNR} &
\textbf{RMSPE} & \textbf{FPR} & \multicolumn{1}{c@{}}{\textbf{FNR}} \\
\hline
Lasso & 1.18 & 0.10 & 0.00 & 2.44 & 0.54 & 0.09 & 2.20 & 0.00 & 0.16 \\
LAD-lasso & 1.13 & 0.05 & 0.00 & 1.15 & 0.07 & 0.00 & 1.27 & 0.18 &
0.00 \\
RLARS & 1.14 & 0.07 & 0.00 & 1.12 & 0.03 & 0.00 & 1.22 & 0.09 & 0.00 \\
Raw sparse LTS & 1.29 & 0.34 & 0.00 & 1.26 & 0.32 & 0.00 & 1.26 & 0.26
& 0.00 \\
Sparse LTS & 1.24 & 0.22 & 0.00 & 1.22 & 0.25 & 0.00 & 1.22 & 0.18 &
0.00 \\
Oracle & 0.82 & & & 0.82 & & & 0.82 & & \\
\hline
\end{tabular*}
\end{table}

\subsubsection{Results for the first sampling scheme}

The simulation results for the first data configuration are displayed in
Table~\ref{tabkhan}. Keep in mind that this configuration is exactly
the same
as in \citet{khan07b}, and that the contamination settings are a
subset of the
ones applied in their paper. In the scenario without contamination, LAD-lasso,
RLARS and lasso show excellent performance with low RMSPE and FPR. The
prediction performance of sparse LTS is good, but it has a larger FPR
than the
other three methods. The reweighting step clearly improves the
estimates, which
is reflected in the lower values for RMSPE and FPR. Furthermore, none
of the
methods suffer from false negatives.

In the case of vertical outliers, the nonrobust lasso is clearly
influenced by
the outliers, reflected in the much higher RMSPE and FPR. RLARS,
LAD-lasso and
sparse LTS, on the other hand, keep their excellent behavior. Sparse
LTS still
has a considerable tendency toward false positives, but the reweighting
step is
a significant improvement over the raw estimator.

When leverage points are introduced in addition to the vertical
outliers, the
performance of RLARS, sparse LTS and LAD-lasso is comparable. The FPR
of RLARS
and LAD-lasso slightly increased, whereas the FPR of sparse LTS slightly
decreased. The LAD-lasso still performs well, and even the lasso
performs better
than in the case of only vertical outliers. This suggests that the leverage
points in this example do not have a bad leverage effect.

%
\begin{figure}

\includegraphics{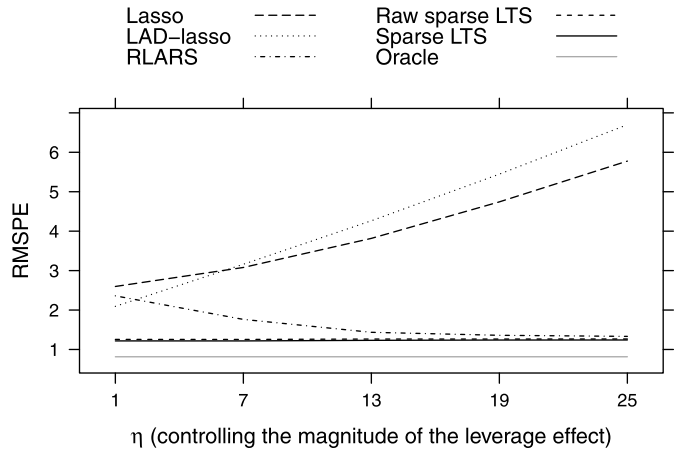}

\caption{Root mean squared prediction error (RMSPE) for the first simulation
scheme, with $n=150$ and $p=50$, and for the fourth contamination setting,
averaged over 500 simulation runs. Lines for raw and reweighted sparse LTS
almost coincide.}
\label{figkhan}
\end{figure}

%
\begin{table}[b]
\tabcolsep=0pt
\caption{Results for the second simulation scheme, with $n=100$ and $p=1000$.
Root mean squared prediction error (RMSPE), the false positive rate
(FPR) and
the false negative rate (FNR), averaged over 500 simulation runs, are reported
for every method}
\label{tabhighdim}
\begin{tabular*}{\textwidth}{@{\extracolsep{\fill}}lccccccccc@{}}
\hline& \multicolumn{3}{c}{\textbf{No contamination}} &\multicolumn{3}{c}{\textbf{Vertical outliers}}
&\multicolumn{3}{c@{}}{\textbf{Leverage points}}
\\[-6pt]
& \multicolumn{3}{c}{\hrulefill} &\multicolumn{3}{c}{\hrulefill} &\multicolumn{3}{c@{}}{\hrulefill} \\
\textbf{Method} & \textbf{RMSPE} & \textbf{FPR} & \textbf{FNR} & \textbf{RMSPE} & \textbf{FPR} & \textbf{FNR} &
\textbf{RMSPE} & \textbf{FPR} & \multicolumn{1}{c@{}}{\textbf{FNR}} \\
\hline
Lasso & 0.62 & 0.00 & 0.00 & 2.56 & 0.08 & 0.16 & 2.53 & 0.00 & 0.71 \\
LAD-lasso & 0.66 & 0.08 & 0.00 & 0.82 & 0.00 & 0.01 & 1.17 & 0.08 &
0.01 \\
RLARS & 0.60 & 0.01 & 0.00 & 0.73 & 0.00 & 0.10 & 0.92 & 0.02 & 0.09 \\
Raw sparse LTS & 0.81 & 0.02 & 0.00 & 0.73 & 0.02 & 0.00 & 0.73 & 0.02
& 0.00 \\
Sparse LTS & 0.74 & 0.01 & 0.00 & 0.69 & 0.01 & 0.00 & 0.71 & 0.02 &
0.00 \\
Oracle & 0.50 & & & 0.50 & & & 0.50 & & \\
\hline
\end{tabular*}
\end{table}

In Figure~\ref{figkhan} the results for the fourth contamination
setting are
shown. The \mbox{RMSPE} is thereby plotted as a function of the parameter
$\eta$. With
increasing $\eta$, the \mbox{RMSPE} of the lasso and the LAD-lasso increases.
\mbox{RLARS} has
a considerably higher \mbox{RMSPE} than sparse LTS for lower values of $\eta
$, but the
\mbox{RMSPE} gradually decreases with increasing $\eta$. However, the RMSPE
of sparse
LTS remains the lowest, thus, it has the best overall performance.

\subsubsection{Results for the second sampling scheme}

Table~\ref{tabhighdim} contains the simulation results for the moderate
high-dimensional data configuration. In the scenario without contamination,
RLARS and the lasso perform best with very low \mbox{RMSPE} and almost perfect
FPR and FNR. Also, the LAD-lasso has excellent prediction performance, followed
by sparse LTS. The LAD-lasso leads to a slightly higher FPR than the other
methods, though. When vertical outliers are added, RLARS still has excellent
prediction performance despite some false negatives. We see that the
sparse LTS
performs best here. In addition, the prediction performance of the nonrobust
lasso already suffers greatly from the vertical outliers. In the
scenario with
additional leverage points, sparse LTS remains stable and is still the
best. For
RLARS, sparsity behavior according to FPR and FNR does not change significantly
either, but there is a small increase in the RMSPE. On the other hand,
LAD-lasso already has a considerably larger RMSPE than sparse LTS, and
again a
slightly higher FPR than the other methods. Furthermore, the lasso is still
highly influenced by the outliers, which is reflected in a very high
FNR and
poor prediction performance.

%
\begin{figure}

\includegraphics{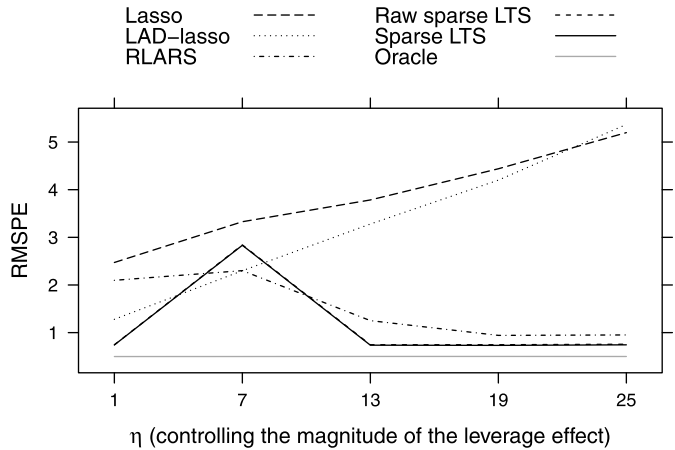}

\caption{Root mean squared prediction error (RMSPE) for the second simulation
scheme, with $n=100$ and $p=1000$, and for the fourth contamination setting,
averaged over 500 simulation runs. Lines for raw and reweighted sparse LTS
almost coincide.}\label{fighighdim}\vspace*{-3pt}
\end{figure}

The results for the fourth contamination setting are presented in
Figure~\ref{fighighdim}. As for the previous simulation scheme, the
RMSPE for
the lasso and the LAD-lasso is increasing with increasing parameter
$\eta$. The
RMSPE for RLARS, however, is gradually decreasing. Sparse LTS shows particularly
interesting behavior: the RMSPE is close to the oracle at first, then
there is a
kink in the curve (with the value of the RMSPE being in between those
for the
LAD-lasso and the lasso), after which the RMSPE returns to low values
close to
the oracle. In any case, for most of the investigated values of $\eta
$, sparse
LTS has the best performance.

\subsubsection{Results for the third sampling scheme}

Table~\ref{tabultrahighdim} contains the simulation results for the more
extreme high-dimensional data configuration. Note that the LAD-lasso
was no
longer computationally feasible with such a large number of variables. In
addition, the number of simulation runs was reduced from 500 to 100 to
lower the
computational effort.

In the case without contamination, the sparse LTS suffers from an efficiency
problem, which is reflected in larger values for RMSPE and FNR than for the
other methods. The lasso and RLARS have considerably better performance
in this
case. With vertical outliers, the RMSPE for the lasso increases
greatly\vadjust{\goodbreak}
due to
many false negatives. Also, RLARS has a larger FNR than sparse LTS,
resulting in
a slightly lower RMSPE for the reweighted version of the latter. When leverage
points are introduced, sparse LTS clearly exhibits the lowest RMSPE and FNR.
Furthermore, the lasso results in a very large FNR.

%
\begin{table}
\def\arraystretch{0.9}
\tabcolsep=0pt
\caption{Results for the third simulation scheme, with $n=100$ and
$p=20\mbox{,}000$.
Root mean squared prediction error (RMSPE), the false positive rate
(FPR) and
the false negative rate (FNR), averaged over 100 simulation runs, are reported
for every method}
\label{tabultrahighdim}
\begin{tabular*}{\textwidth}{@{\extracolsep{\fill}}lccccccccc@{}}
\hline& \multicolumn{3}{c}{\textbf{No contamination}} &\multicolumn{3}{c}{\textbf{Vertical outliers}}
&\multicolumn{3}{c@{}}{\textbf{Leverage points}}
\\[-6pt]
& \multicolumn{3}{c}{\hrulefill} &\multicolumn{3}{c}{\hrulefill} &\multicolumn{3}{c@{}}{\hrulefill} \\
\textbf{Method} & \textbf{RMSPE} & \textbf{FPR} & \textbf{FNR} & \textbf{RMSPE} & \textbf{FPR} & \textbf{FNR} &
\textbf{RMSPE} & \textbf{FPR} & \multicolumn{1}{c@{}}{\textbf{FNR}} \\
\hline
Lasso & 1.43 & 0.000 & 0.00 & 5.19 & 0.004 & 0.49 & 5.57 & 0.000 & 0.83
\\
RLARS & 1.54 & 0.001 & 0.00 & 2.53 & 0.000 & 0.38 & 3.34 & 0.001 & 0.45
\\
Raw sparse LTS & 3.00 & 0.001 & 0.19 & 2.59 & 0.002 & 0.11 & 2.59 &
0.002 & 0.10 \\
Sparse LTS & 2.88 & 0.001 & 0.16 & 2.49 & 0.002 & 0.10 & 2.57 & 0.002 &
0.09 \\
Oracle & 1.00 & & & 1.00 & & & 1.00 & & \\
\hline
\end{tabular*}   \vspace*{-3pt}
\end{table}

%
\begin{figure}[b]\vspace*{-3pt}

\includegraphics{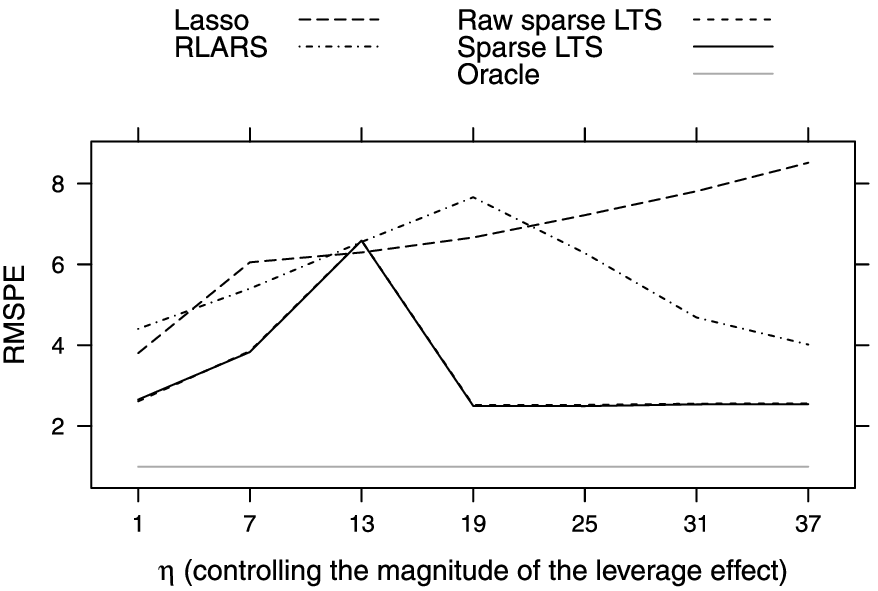}

\caption{Root mean squared prediction error (RMSPE) for the third simulation
scheme, with $n=100$ and $p=20\mbox{,}000$, and for the fourth contamination setting,
averaged over 100 simulation runs. Lines for raw and reweighted sparse LTS
almost coincide.}
\label{figultrahighdim}
\end{figure}

Figure~\ref{figultrahighdim} shows the results for the fourth contamination
setting.
Most interestingly, the RMSPE of RLARS in this case keeps increasing in the
beginning and even goes above the one of the lasso, before dropping dropping
continuously in the remaining steps. Sparse LTS again shows a kink in
the curve
for the RMSPE, but clearly performs best.\vspace*{-3pt}

\subsubsection{Summary of the simulation results}
Sparse LTS shows the best overall performance in this simulation study,
if the
reweighted version is taken. Concerning\vadjust{\goodbreak} the other investigated methods, RLARS
also performs well, but suffers sometimes from an increased percentage
of false
negatives under contamination. It is also confirmed that the lasso is
not robust
to outliers. The LAD-lasso still sustains vertical outliers, but is not robust
against bad leverage points.


\section{NCI-60 cancer cell panel} \label{secex}
In this section the sparse LTS estimator is compared to the competing methods
in an application to the cancer cell panel of the National Cancer Institute.
It consists of data on 60 human cancer cell lines and can be downloaded
via the
web application \proglang{CellMiner}
(\url{http://discover.nci.nih.gov/cellminer/}). We regress protein
expression on
gene expression data. The gene expression data were obtained with an Affymetrix
HG-U133A chip and normalized with the GCRMA method, resulting in a set
of $p =
22\mbox{,}283$ predictors. The protein expressions based on 162 antibodies were
acquired via reverse-phase protein lysate arrays and $\log_{2}$
transformed. One
observation had to be removed since all values were missing in the gene
expression data, reducing the number of observations to $n = 59$. More details
on how the data were obtained can be found in \citet{shankavaram07}.
Furthermore, \citet{lee11} also use this data for regression analysis, but
consider only nonrobust methods.
They obtain models that still consist of several hundred to several thousand
predictors and are thus difficult to interpret.

Similar to \citet{lee11}, we first order the protein expression variables
according to their scale, but use the MAD (median absolute deviation
from the
median, multiplied with the consistency factor 1.4826) as a scale estimator
instead of the standard deviation. We show the results for the protein
expressions based on the KRT18 antibody, which constitutes the variable
with the
largest MAD, serving as one dependent variable. Hence, our response variable
measures the expression levels of the protein \textit{keratin 18},
which is known
to be persistently expressed in carcinomas [\citet{oshima96}].
We compare raw and reweighted sparse LTS with 25\% trimming, lasso and RLARS.
As in the simulation study, the LAD-lasso could not be computed for
such a large
$p$. The optimal models are selected via BIC as discussed in
Section~\ref{seclambda}. The raw sparse LTS estimator thereby results
in a
model with 32 genes. In the reweighting step, one more observation is
added to
the best subset found by the raw estimator, yielding a model with 33
genes for
reweighted sparse LTS (thus also one more gene is selected compared to
the raw
estimator). The lasso model is somewhat larger with 52 genes, whereas
the RLARS
model is somewhat smaller with 18 genes.

Sparse LTS and the lasso have three selected genes in common, one of
which is
KRT8. The product of this gene, the protein \textit{keratin 8},
typically forms an
intermediate filament with keratin 18 such that their expression levels are
closely linked [e.g., \citet{owens03}]. However, the larger model of
the lasso
is much more difficult to interpret. Two of the genes selected by the
lasso are
not even recorded in the \proglang{Gene} database [\citet{maglott05}]
of the
National Center for Biotechnology Information (NCBI). The sparse LTS
model is
considerably smaller and easier to interpret. For instance, the gene expression
level of MSLN, whose product \textit{mesothelin} is overexpressed in
various forms
of cancer [\citet{hassan04}], has a positive effect on the protein expression
level of keratin 18.

%
\begin{table}
\tablewidth=200pt
\caption{Root trimmed mean squared prediction error (RTMSPE) for protein
expressions based on the KRT18 antibody (NCI-60 cancer cell panel data),
computed from leave-one-out cross-validation}
\label{tabnci60-RTMSPE}
\begin{tabular*}{200pt}{@{\extracolsep{\fill}}lc@{}}
\hline
\textbf{Method} & \textbf{RTMSPE} \\
\hline
Lasso & 1.058 \\
RLARS & 0.936 \\
Raw sparse LTS & 0.727 \\
Sparse LTS & 0.721 \\
\hline
\end{tabular*}
\end{table}

Concerning prediction performance, the root trimmed mean squared prediction
error (RTMSPE) is computed as in \eqref{RTMSPE} via leave-one-out
cross-validation (so $k = n$).
Table~\ref{tabnci60-RTMSPE} reports the RTMSPE for the considered methods.
Sparse LTS clearly shows the smallest RTMSPE, followed by RLARS and the lasso.
In addition, sparse LTS detects 13 observations as outliers, showing
the need
for a robust procedure.
Further analysis revealed that including those 13 observations changes the
correlation structure of the predictor variables with the response.
Consequently, the order in which the genes are added to the model by
the lasso
algorithm on the full sample is completely different from the order on
the best
subset found by sparse LTS. Leaving out those 13 observations therefore yields
more reliable results for the majority of the cancer cell lines.

It is also worth noting that the models still contain a rather large
number of
variables given the small number of observations. For the lasso, it is well
known that it tends to select many noise variables in high dimensions
since the
same penalty is applied on all variables. \citet{meinshausen07} therefore
proposed a relaxation of the penalty for the selected variables of an initial
lasso fit. Adding such a relaxation step to the sparse LTS procedure
may thus be
beneficial for large $p$ and is considered for future work.


%
\begin{figure}

\includegraphics{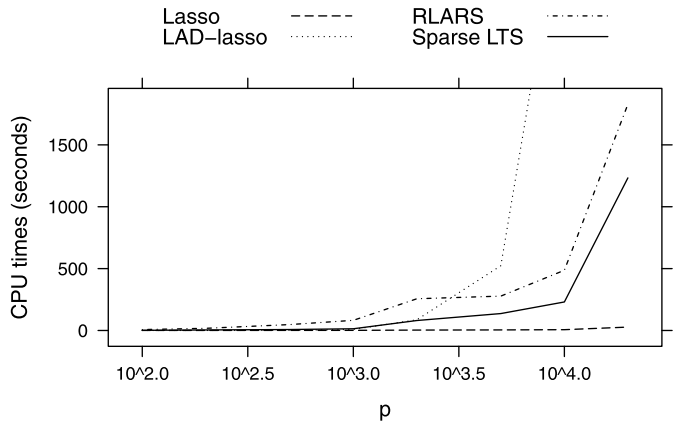}

\caption{CPU times (in seconds) for $n = 100$ and varying $p$,
averaged over 10
runs.}\label{figcpu}
\end{figure}

\section{Computational details and CPU times} \label{seccpu}
All computations are carried out in \proglang{R} version 2.14.0 [\citet{RDev}]
using the packages \pkg{robustHD} [\citet{robustHD}] for sparse LTS
and RLARS,
\pkg{quantreg} [\citet{quantreg}] for the LAD-lasso and \pkg{lars}
[\citet{lars}]
for the lasso. Most of sparse\vadjust{\goodbreak} LTS is thereby implemented in \proglang{C++},
while RLARS is an optimized version of the~\proglang{R} code by \citet
{khan07b}.
Optimization of the RLARS code was necessary since the original code
builds a $p
\times p$ matrix of robust correlations, which is not computationally feasible
for very large $p$. The optimized version only stores an $q \times p$ matrix,
where $q$ is the number of sequenced variables. Furthermore, the robust
correlations are computed with \proglang{C++} rather than \proglang{R}.

Since computation time is an important practical consideration,
Figure~\ref{figcpu} displays computation times of lasso, LAD-lasso,
RLARS and
sparse LTS in seconds. Note that those are average times over 10 runs
based on
simulated data with $n = 100$ and varying dimension $p$, obtained on an Intel
Xeon X5670 machine. For sparse LTS and the LAD-lasso, the reported CPU
times are
averages over a grid of five values for $\lambda$. RLARS is a hybrid procedure,
thus, we only report the CPU times for obtaining the sequence of
predictors, but
not for fitting the models along the sequence.


As expected, the computation time of the nonrobust lasso remains very
low for
increasing $p$. Sparse LTS is still reasonably fast up to $p \approx10\mbox{,}000$,
but computation time is a considerable factor if $p$ is much larger
than that.
However, sparse LTS remains faster than obtaining the RLARS sequence. A further
advantage of the subsampling algorithm of sparse LTS is that it can
easily be
parallelized to reduce computation time on modern multicore computers,
which is
future work.


\section{Conclusions and discussion}
\label{secconcl}
Least trimmed squares (LTS) is a robust regression method frequently
used in
practice. Nevertheless, it does not allow for sparse model estimates
and cannot
be applied to high-dimensional data with $p > n$. This paper introduced the
sparse LTS estimator, which overcomes these two issues simultaneously
by adding
an $L_{1}$ penalty to the LTS objective function.\vadjust{\goodbreak} Simulation results
and a real
data application to protein and gene expression data of the NCI-60
cancer cell
panel illustrated the excellent performance of sparse LTS and showed
that it
performs as well or better than robust variable selection methods such as
RLARS. In addition, an advantage of sparse LTS over algorithmic
procedures such
as RLARS is that the objective function allows for theoretical
investigation of
its statistical properties. As such, we could derive the breakdown
point of the
sparse LTS estimator. However, it should be noted that efficiency is an issue
with sparse LTS. A reweighting step can thereby lead to a substantial
improvement in efficiency, as shown in the simulation study.

In the paper, an $L_1$ penalization was imposed on the regression
parameter, as
for the lasso. Other choices for the penalty are possible. For example, an
$L_{2}$ penalty leads to ridge regression. A robust version of ridge regression
was recently proposed by \citet{maronna11}, using $L_{2}$ penalized
MM-estimators. Even though the resulting estimates are not sparse, prediction
accuracy is improved by shrinking the coefficients, and the
computational issues
with high-dimensional robust estimators are overcome due to the regularization.
Another possible choice for the penalty function is the smoothly clipped
absolute deviation penalty (SCAD) proposed by \citet{fan01}. It
satisfies the
mathematical conditions for sparsity but results in a more difficult
optimization problem than the lasso. Still, a robust version of SCAD
can be
obtained by optimizing the associated objective function over trimmed samples
instead of over the full sample.

There are several other open questions that we leave for future
research. For
instance, we did not provide any asymptotics for sparse LTS, as was,
for example,
done for penalized M-estimators in \citet{germain09}. Potentially,
sparse LTS
could be used as an initial estimator for computing penalized M-estimators.

All in all, the results presented in this paper suggest that sparse LTS
is a
valuable addition to the statistics researcher's toolbox. The sparse LTS
estimator has an intuitively appealing definition and is related to the popular
least trimmed squares estimator of robust regression. It performs model
selection, outlier detection and robust estimation simultaneously, and is
applicable if the dimension is larger than the sample size.


\begin{appendix}

\section*{Appendix: Proof of breakdown point}\label{app}

\begin{pf*}{Proof of Theorem~\ref{thmfbp}} In this proof the $L_1$
norm of a
vector $\bolds{\beta}$ is denoted as $\| \bolds{\beta} \|_1$ and
the Euclidean
norm as $\| \bolds{\beta} \|_2$. Since these norms are topologically
equivalent,
there exists a constant $c_1>0$ such that $\| \bolds{\beta} \|_1 \geq
c_1 \|
\bolds{\beta} \|_2$ for all vectors~$\bolds{\beta}$. The proof is
split into two
parts.

First, we prove that $\varepsilon^{*}(\hat{\bolds{\beta}}; \mymat
{Z}) \geq
\frac{n-h+1}{n}$. Replace the last $m \leq n-h$ observations,
resulting in the
contaminated sample $\tilde{\mymat Z}$. Then there are still $n-m \geq
h$ good
observations in $\tilde{\mymat Z}$. Let $M_{y} = \max_{1 \leq i \leq
n} |y_{i}|$
and $M_{x_{1}} = \max_{1 \leq i \leq n} |x_{i1}|$. For the case $\beta_{j} =
0$, $j = 1, \ldots, p$, the value of the objective function is given by
%
\[
Q(\myvec{0}) = \sum_{i=1}^{h}
\bigl( \bolds{\rho}(\bolds{\tilde y}) \bigr)_{i:n} \leq\sum
_{i=1}^{h} \bigl( \bolds{\rho}(\myvec{y})
\bigr)_{i:n} \leq h \rho(M_{y}).
\]
Now consider any $\bolds{\beta}$ with $\| \bolds{\beta} \|_{2} \geq
M := (h
\rho(M_{y}) + 1)/(\lambda c_{1})$. For the value of the objective
function, it
holds that 
\[
Q(\bolds{\beta}) \geq\lambda\| \bolds{\beta} \|_{1} \geq
\lambda c_{1} \| \bolds{\beta} \|_{2} \geq h
\rho(M_{y}) + 1 > Q(\myvec{0}).
\]
Since $Q(\hat{\bolds{\beta}}) \leq Q(0)$, we conclude that $\|
\hat{\bolds{\beta}}(\tilde{\mymat Z}) \|_{2} \leq M$, where $M$
does not depend
on the outliers. This concludes the first part of the proof.

Second, we prove that $\varepsilon^{*}(\hat{\bolds{\beta}}; \mymat
{Z}) \leq
\frac{n-h+1}{n}$. Move the last $m = n-h+1$ observations of $\mymat
{Z}$ to
the position $\myobs{z}(\gamma, \tau) = (\myobs{x}(\tau)',
y(\gamma, \tau))' =
((\tau, 0, \ldots, 0), \gamma\tau)'$ with $\gamma, \tau> 0$, and denote
$\mymat{Z}_{\gamma, \tau}$ the resulting contaminated sample. Assume
that there
exists a constant M such that
%
\renewcommand{\theequation}{A.\arabic{equation}}
\setcounter{equation}{0}
\begin{equation}
\label{eqassumption} \sup_{\tau, \gamma} \bigl\| \hat{\bolds{\beta}}(
\mymat{Z}_{\gamma,
\tau} ) \bigr\|_{2} \leq M,
\end{equation}
that is, there is no breakdown. We will show that this leads to a contradiction.

Let $\bolds{\beta}_{\gamma} = (\gamma, 0, \ldots, 0)' \in\mathbb
{R}^{p}$ with
$\gamma= M+2$ and define $\tau> 0$ such that $\rho(\tau) \geq\max
(h-m, 0)
\rho(M_{y} + \gamma M_{x_{1}}) + h \lambda\gamma+ 1$. Note that
$\tau$ is
always well defined due to the assumptions on $\rho$, in particular, since
$\rho(\infty) = \infty$. Then the objective function is given by
%
\[
Q(\bolds{\beta}_{\gamma}) = \cases{
\displaystyle\sum_{i=1}^{h-m} \bigl( \bolds{\rho}(y -
\mymat{X}\bolds{\beta }_{\gamma}) \bigr)_{i:(n-m)} + h \lambda|
\gamma|, &\quad  $\mbox{if } h > m,$\vspace*{2pt}\cr
h \lambda| \gamma|, & \quad $\mathrm{else},$}
\]
since the residuals with respect to the outliers are all zero. Hence,
%
\begin{equation}
\label{eq1} Q(\bolds{\beta}_{\gamma}) \leq\max(h-m, 0)
\rho(M_{y} + \gamma M_{x_{1}}) + h \lambda\gamma\leq\rho(\tau) -
1.
\end{equation}

Furthermore, for $\bolds{\beta} = (\beta_{1}, \ldots, \beta_{p})'$
with $\|
\bolds{\beta} \|_{2} \leq\gamma-1$ we have
%
\[
Q(\bolds{\beta}) \geq\rho(\gamma\tau- \tau\beta_{1}),
\]
since at least one outlier will be in the set of the smallest $h$
residuals. Now
$\beta_{1} \leq\| \bolds{\beta} \|_{2} \leq\gamma-1$, so that
%
\begin{equation}
\label{eq2} Q(\bolds{\beta}) \geq\rho\bigl(\tau(\gamma- \beta_{1})
\bigr) \geq\rho (\tau),
\end{equation}
since $\rho$ is nondecreasing.

Combining (\ref{eq1}) and (\ref{eq2}) leads to
%
\[
\bigl\| \hat{\bolds{\beta}}(\mymat{Z}_{\gamma, \tau}) \bigr\|_{2} \geq
\gamma- 1 = M + 1,
\]
which contradicts the assumption (\ref{eqassumption}). Hence, there
is breakdown.
\end{pf*}
\end{appendix}

\section*{Acknowledgments}
We would like to thank the Editor and two anonymous referees for their
constructive remarks that led to an improvement of the paper.


%

%

%


\printaddresses

\end{document}